\documentclass{aa}
\usepackage{rotating}
\usepackage{graphicx}
\usepackage{txfonts}
\begin{document}

\title{Early-type stars in the core of the young open cluster Westerlund\,2\thanks{Based on observations collected at the Cerro Tololo Interamerican Observatory and at the European Southern Observatory (La Silla, Chile)}\thanks{Tables 2, 3 and 4 are only available in electronic form at the CDS via anonymous ftp to cdsarc.u-strasbg.fr (130.79.128.5) or via http://cdsweb.u-strasbg.fr/cgi-bin/qcat?J/A+A/}}
\author{G.\ Rauw\inst{1}\fnmsep\thanks{Research Associate FNRS (Belgium)} \and J.\ Manfroid\inst{1}\fnmsep\thanks{Research Director FNRS (Belgium)} \and E.\ Gosset\inst{1}\fnmsep$^{\star\star}$ \and Y.\ Naz\'e\inst{1}\fnmsep\thanks{Postdoctoral Researcher FNRS (Belgium)} \and H.\ Sana\inst{2,1} \and\\ M.\ De Becker\inst{1} \and C.\ Foellmi\inst{3,2} \and A.F.J.\ Moffat\inst{4}}
\offprints{G.\ Rauw}
\mail{rauw@astro.ulg.ac.be}
\institute{Institut d'Astrophysique et de G\'eophysique, Universit\'e de Li\`ege, All\'ee du 6 Ao\^ut, B\^at B5c, 4000 Li\`ege, Belgium \and
European Southern Observatory, Alonso de Cordova 3107, 19 Vitacura, Santiago, Chile \and
Observatoire de Grenoble, 414, Rue de la Piscine, BP 53, 38041 Grenoble, France
\and D\'epartement de Physique, Universit\'e de Montr\'eal, QC, H3C 3J7, and Observatoire du Mont M\'egantic, Canada}
\date{Received date / Accepted date}
\abstract{}{The properties of the early-type stars in the core of the \object{Westerlund\,2} cluster are examined in order to establish a link between the cluster and the very massive Wolf-Rayet binary \object{WR\,20a} as well as the H\,{\sc ii} complex \object{RCW\,49}.}{Photometric monitoring as well as spectroscopic observations of Westerlund\,2 are used to search for light variability and to establish the spectral types of the early-type stars in the cluster core.}{The first light curves of the eclipsing binary WR\,20a in $B$ and $V$ filters are analysed and a distance of 8\,kpc is inferred. Three additional eclipsing binaries, which are probable late O or early B-type cluster members, are discovered, but none of the known early O-type stars in the cluster displays significant photometric variability above 1\% at the 1-$\sigma$ level. The twelve brightest O-type stars are found to have spectral types between O3 and O6.5, significantly earlier than previously thought.}{The distance of the early-type stars in Westerlund\,2 is established to be in excellent agreement with the distance of WR\,20a, indicating that WR\,20a actually belongs to the cluster. Our best estimate of the cluster distance thus amounts to $8.0 \pm 1.4$\,kpc. Despite the earlier spectral types, the currently known population of early-type stars in Westerlund\,2 does not provide enough ionizing photons to account for the radio emission of the RCW\,49 complex. This suggests that there might still exist a number of embedded early O-stars in RCW\,49.}
\maketitle
\section{Introduction}
The open cluster Westerlund\,2 (Westerlund \cite{Westerlund}) lies in a blowout region of the giant H\,{\sc ii} complex RCW\,49 (Rodgers et al.\ \cite{RCW}). The stellar winds and ionizing radiation of the early-type stars in Westerlund\,2 have evacuated the dust in the central part of RCW\,49 and filled the cavity with very hot low density gas. Unfortunately, so far, our knowledge of the stellar content of Westerlund\,2 was rather fragmentary. Photometric studies were performed by Moffat \& Vogt (\cite{MV}), Moffat et al.\ (\cite{MSP}) and Carraro \& Munari (\cite{CM}). To date, the only spectroscopic study of the cluster was presented by Moffat et al.\ (\cite{MSP}) who obtained low resolution spectra of six O-stars in Westerlund\,2 and classified them as O6-7:V objects (although, as indicated by the colon, these spectral classifications were rather uncertain). 

Recently, interest in the stellar population of Westerlund\,2 was triggered by two independent observational studies. On the one hand, RCW\,49 was observed with the Infrared Array Camera (IRAC) aboard {\it Spitzer} in the framework of the GLIMPSE legacy survey (Whitney et al.\ \cite{Whitney}, Churchwell et al.\ \cite{Churchwell04}, Churchwell \cite{Churchwell05}). These observations revealed strong evidence for ongoing star formation activity and underlined the need for a detailed study of the properties of the early-type stars in the cluster core and their impact on the surrounding nebula. On the other hand, WR\,20a - one of the two Wolf-Rayet stars in RCW\,49 - was found to be a very massive eclipsing binary consisting of two WN6ha stars with individual masses of about 80\,M$_{\odot}$ (Rauw et al.\ \cite{Rauw04,Rauw05}, Bonanos et al.\ \cite{bonanos}). Since previous spectroscopic studies revealed only one O6: and six O7: stars, in the cluster core (Moffat et al.\ \cite{MSP}), the existence of a pair of stars that massive was somewhat unexpected. These results therefore called for a re-investigation of the population of early-type stars in Westerlund\,2.
\begin{figure*}[t!]
\begin{minipage}{8.0cm}
%\resizebox{8.0cm}{!}{\includegraphics{6495f1a.ps}}
\end{minipage}
\hfill
%\hspace*{-2cm}
\begin{minipage}{9.0cm}
%\resizebox{9.0cm}{!}{\includegraphics{6495f1b.ps}}
\end{minipage}
\caption{{\it Left panel}: average image (combining the best $B$ and $V$ filter observations, i.e.\ those with a seeing of less than 1.25\,arcsec) of the region around Westerlund\,2 as observed with the ANDICAM instrument. {\it Right panel}: zoom on the central part of the cluster. The labels correspond to the numbering scheme introduced by Moffat et al.\ (\cite{MSP}) and identify those stars for which we obtained spectra. Note that MSP\,158 and 218 are late-type foreground stars that are unrelated to the cluster (see Sect.\,\ref{latety}). The boxes indicate the three newly discovered eclipsing binary systems discussed below.\label{image}}
\end{figure*}

% In this paper, we present the results of a photometric monitoring as well as spectroscopic observations of the core of this cluster.  
\section{Observations}
\subsection{Photometry\label{photom}}
Photometric monitoring of the core of Westerlund\,2 was performed in service mode with the ANDICAM instrument at the 1.3\,m telescope at CTIO operated by the SMARTS\footnote{Small and Moderate Aperture Research Telescope System} consortium. The primary goal of this campaign was the study of the light curves of early-type stars and more specifically of the very massive eclipsing binary WR\,20a. Whenever possible, the cluster was observed twice per night (separated by several hours) through the $B$ and $V$ filters of the ANDICAM instrument. Each pointing consisted of integration times of $2 \times 200$ and $2 \times 75$\,s for the $B$ and $V$ filters respectively. The detector was a Fairchild 447 CCD of $2048 \times 2048$ pixels. In the standard mode of the instrument the pixels are binned by $2 \times 2$. Each binned pixel corresponds to 0.37\,\arcsec on the sky and the total field of view of the instrument is about $6 \times 6$\,arcmin$^2$, well suited for monitoring simultaneously most of the early-type stars in the central part of Westerlund\,2 (see Fig.\,\ref{image}).
 
Our campaign covered 45 consecutive nights from 3 December 2004 to 17 January 2005. The median seeing of our data is 1.4\,arcsec. During photometric nights, observations of Landolt (\cite{Landolt}) standard star fields were also performed. All the data were pipeline processed (bias subtraction and division by a normalized flat field) by the SMARTS consortium at Yale before they were delivered to us. 

The photometric reduction was done in three steps. First, instrumental magnitudes were extracted with DAOPHOT (Stetson \cite{Stetson})
under {\sc iraf}\footnote{{\sc iraf} is distributed by the National Optical Astronomy Observatories, which are operated by the Association of Universities for Research in Astronomy, Inc., under cooperative agreement with the National Science Foundation.} using aperture and PSF-fitting techniques. To avoid possible flat-fielding errors (Manfroid \cite{Jean95}, Manfroid et al.\ \cite{Manfroid}) the atmospheric extinction and zero-point were determined for the best photometric nights, using only the central part of the fields, by a  multi-night, multi-star, multi-filter method as described by Manfroid (\cite{Jean93}). The implementation of this reduction procedure allows the 
construction of a consistent natural system, which contains the extra-atmospheric instrumental magnitudes of all constant stars included in the computation procedure. Stars from all observed fields were included in the building of the natural system. To prevent adverse PSF effects, this procedure was carried out with the widest aperture photometry and restricted to relatively bright, isolated stars. The external calibration is based on two Landolt (\cite{Landolt}) standard fields, T\,Phe and Rubin\,149, revisited by Stetson \footnote{http://cadcwww.hia.nrc.ca/standards/}. These fields were observed at air masses ranging between 1.0 and 2.7. The {\tt rms} error on the extinction coefficients of each night is 0.018. The error on the zero points is 0.028\,mag, whilst the {\tt rms} deviation of the mean magnitudes of the standard stars is slightly less than 0.003\,mag.
Finally, an extended set of constant stars was built to provide a calibration for all the frames, including those of relatively poor quality, and over the whole field. This procedure (Manfroid \cite{Jean95}) yields robust differential data. As the centering of the fields was quite consistent over the whole
data set, inaccuracies in the overall flat-field calibration could not be evaluated, but they can be tolerated since they only consist of a systematic space-dependent shift and do not affect the variable component of the light curves. 

Comparing our PSF photometry to the values reported by Moffat et al.\ (\cite{MSP}) - calibrated by the photomultiplier aperture photometry of Moffat \& Vogt (\cite{MV}) - for 59 objects in common (excluding a few outliers) and Carraro \& Munari (\cite{CM}) for 82 objects in common, we obtain:

$(B - V)_{MSP} = (0.860 \pm 0.024) (B - V) + (0.089 \pm 0.037)$

$(B - V)_{CM} = (0.958 \pm 0.016) (B - V) + (0.025 \pm 0.020)$

$V_{MSP} - V = (0.052 \pm 0.029) (B - V) - (0.086 \pm 0.044)$

$V_{CM} - V = (0.062 \pm 0.024) (B - V) - (0.183 \pm 0.029)$\\
where magnitudes and colours without an index refer to our data. As noted by Carraro \& Munari (\cite{CM}), the colour effects in the Moffat et al.\ (\cite{MSP}) data are large. Whilst our $V$ zero-point agrees with the one of Moffat et al., the discrepancy with the zero-point of Carraro \& Munari cannot be easily understood since most of the 82 stars in common are not affected by blends. 
\subsection{Spectroscopy}
For the spectroscopic analysis, we adopt the MSP numbering convention introduced by Moffat et al.\ (\cite{MSP}, see also Fig.\,\ref{image}). Medium-resolution long-slit spectroscopy of fourteen of the brightest stars in the central part of the Westerlund\,2 cluster was obtained with the EMMI instrument at the ESO New Technology Telescope (NTT) on 27 March 2005. We excluded star MSP\,91 for which Moffat et al.\ (\cite{MSP}) obtained a spectral type G0\,V-III as well as MSP\,273, another suspected non-member which lies rather far away from the cluster core. All seven O-type stars observed by Moffat et al.\ (\cite{MSP}) were included among our targets. Finally, MSP\,168 was included in our target list, but could not be observed due to a lack of time.

The red arm of EMMI was used in the REMD mode with grating \#6 (1200 grooves\,mm$^{-1}$). The useful wavelength domain ranged from 4200 to 4950\,\AA. The spectral resolution of the instrument configuration as measured from the {\tt FWHM} of the lines in the ThAr calibration exposures was about 1\,\AA. Exposure times ranged from 20\,min to 1\,hour. The spectra were bias-subtracted, flat-fielded, sky-subtracted, wavelength-calibrated and normalized using the {\sc midas} software developed at ESO. The S/N ratios of the calibrated spectra ranged between 60 and 150 for the faintest and brightest sources respectively, with an average of $\sim 100$ for a $V = 14.5$ star. Since the observations were taken only two days after full moon, the level of the background sky was rather high. Although we extracted the sky spectrum as close as possible to the source, some residual sky features are seen in the final spectra of the faintest source that we observed under somewhat poorer conditions (MSP\,263).

\section{Spectral classifications}
\subsection{Late-type foreground stars \label{latety}}
\begin{figure}[htb]
\resizebox{8.5cm}{!}{\includegraphics{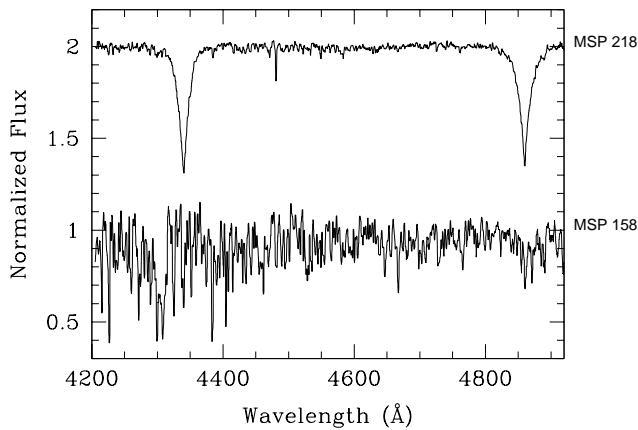}}
\caption{Medium resolution spectra of the stars MSP\,158 and 218 in the field of Westerlund\,2. Both stars are likely foreground objects unrelated to the cluster.\label{latetype}}
\end{figure}

Two stars (\object{MSP\,158} and 218) have spectral types significantly later than B0 (see Fig.\,\ref{latetype}). In order to establish the MK spectral types of these two objects, we make use of the digital spectral classification atlas compiled by R.O.\ Gray and available on the web\footnote{ http://nedwww.ipac.caltech.edu/level5/Gray/frames.html}.  

The spectrum of \object{MSP\,218} (10:24:07.7, $-57$:45:41, J2000) is dominated by strong and broad H\,{\sc i} Balmer lines. Weak absorption lines of He\,{\sc i} $\lambda\lambda$\,4388, 4471, Mg\,{\sc ii} $\lambda$\,4481 and Si\,{\sc iii} $\lambda$\,4552 are also seen. The He\,{\sc i} $\lambda$\,4471/Mg\,{\sc ii} $\lambda$\,4481 intensity ratio indicates a spectral type B8-A0. The appearance of some weak metallic lines favours an A0 classification, whilst the width of the Balmer lines suggests a giant luminosity class. We therefore adopt an A0\,III spectral type. Our photometric data yield $V = 14.13 \pm 0.03$ and $B - V = 0.34 \pm 0.04$. If the intrinsic colours and absolute magnitudes of MSP\,218 are those of a typical A0\,III star (Schmidt-Kaler \cite{SK}), then $A_V \sim 1.15$ and the star must be located at a distance of about 4.0\,kpc.  

The spectrum of MSP\,158 (10:23:59.5, $-57$:45:23, J2000) shows a prominent G-band as well as a strong Ca\,{\sc i} $\lambda$\,4227 line indicating a late (G-K) spectral type. The intensity ratios between the Fe\,{\sc i} $\lambda$\,4325 and H$\gamma$ lines as well as the ratio between Cr\,{\sc i} $\lambda$\,4254 and the neighbouring Fe\,{\sc i} $\lambda\lambda$\,4250, 4260 lines suggest a spectral type G8. The Y\,{\sc ii} $\lambda$\,4376/Fe\,{\sc i} $\lambda$\,4383 line ratio and the CN bandhead at 4216\,\AA\ indicate a giant luminosity class. We thus assign a G8\,III spectral type. From our photometric data, we infer $V = 13.03 \pm 0.03$ and $B - V = 1.09 \pm 0.05$. Assuming magnitudes and colours typical for a G8\,III star (Schmidt-Kaler \cite{SK}), we derive a reddening of $A_V = 0.47$ and a distance of about 2.25\,kpc.

Both stars are foreground objects unrelated to the Westerlund\,2 cluster, as suspected by Moffat et al.\ (\cite{MSP}) based on their $UBV$ photometry. This is also confirmed by the absence of the strong diffuse interstellar band at 4430\,\AA\ which is prominent in the spectra of the O-type members of the cluster (see Fig.\,\ref{Otype}). As we will show below, the O-stars of Westerlund\,2 are subject to substantially larger reddening ($A_V$ from about 4.8 to 5.8\,mag) and their distance is significantly larger. Comparing the reddening of the G8\,III, A0\,III and the O-type stars (see below), we note that the $A_V$ of the cluster members exceeds the value estimated from a crude empirical linear $A_V(d)$ relation derived from the foreground stars. Though such a linear relation provides only a first approximation, the result suggests that much of the reddening that affects the cluster could be local (e.g.\ due to material from the molecular cloud out of which the cluster has formed). 

\begin{table*}[ht!]
\caption{Spectral classifications of the O-type stars in Westerlund\,2. The first and fourth columns yield the number of the star in the numbering scheme of Moffat et al.\ (\cite{MSP}) as well as the spectral type proposed by these authors. The fifth and sixth columns provide our newly derived spectral types, determined using the Conti (\cite{Conti}) + Mathys (\cite{Mathys}) and Walborn \& Fitzpatrick (\cite{WF}) classification criteria. Columns 7, 8 and 9 respectively yield our mean $V$ magnitude, the $B-V$ colour index and the extinction derived assuming the intrinsic colours from Martins \& Plez (\cite{MP}) and $R_V = 3.1$.
The individual spectro-photometric distance moduli based on the calibration of Martins \& Plez (\cite{MP}) are listed in column 10. Column 11 provides the effective temperature as derived from our spectral classification and the calibration of Martins et al.\ (\cite{Martins}). Finally, the bolometric luminosities given in the last column were computed for an adopted distance modulus of $DM = 14.52 \pm 0.38$ ($d = (8.0 \pm 1.4)$\,kpc, see Sect.\,\ref{distance}). \label{types}}
\begin{tabular}{c c c c c c c c c c l c}
\hline
Star \# & RA & DEC & \multicolumn{3}{c}{Spectral type} & $V$ & $B-V$ & $A_V$ & $DM$ & \multicolumn{1}{c}{T$_{\rm eff}$ (K) } & $\log{L_{\rm bol}/L_{\odot}}$\\
\cline{4-6}
& (J2000) & (J2000) & MSP & \multicolumn{2}{c}{This work} & & & & & & \\
\cline{5-6}
& & & & Conti crit. & W\&F crit. & & & & & \\
\hline
 \object{MSP\,18} & 10:24:02.4 & $-57$:44:35 & O7:V & O5.5\,V-III & O5\,V-III((f)) & 12.88 & 1.23 & 4.68 & 13.74 & $40100 \pm 800$  & $5.89 \pm 0.16$ \\
\object{MSP\,151} & 10:24:00.5 & $-57$:45:24 & O6:V & O7\,III     & O6\,III        & 14.33 & 1.36 & 5.05 & 14.97 & $37100 \pm 1000$ & $5.37 \pm 0.17$ \\
\object{MSP\,157} & 10:24:00.8 & $-57$:45:26 & O6:V & O5.5\,V     & O6.5\,V        & 14.33 & 1.31 & 4.93 & 14.39 & $38900 \pm 1000$ & $5.37 \pm 0.17$ \\
\object{MSP\,167} & 10:24:02.1 & $-57$:45:28 & O7:V & O6\,III     & O6\,III        & 15.18 & 1.42 & 5.24 & 15.71 & $38200 \pm 500$  & $5.14 \pm 0.16$ \\
\object{MSP\,171} & 10:24:04.9 & $-57$:45:28 &    & O5\,V       & O4-5\,V        & 14.47 & 1.45 & 5.36 & 14.53 & $41900 \pm 1000$ & $5.58 \pm 0.17$ \\
\object{MSP\,175} & 10:24:01.4 & $-57$:45:30 &    & O5.5\,V-III & O6\,V-III      & 13.98 & 1.29 & 4.87 & 14.55 & $39000 \pm 800$  & $5.49 \pm 0.16$ \\
\object{MSP\,182} & 10:23:56.2 & $-57$:45:30 &   & O4\,V-III   & O4\,V-III((f)) & 14.45 & 1.32 & 4.96 & 15.30 & $42600 \pm 1000$ & $5.45 \pm 0.16$ \\
\object{MSP\,183} & 10:24:02.4 & $-57$:45:31 & O7:V & O4\,V       & O3\,V((f))     & 13.57 & 1.50 & 5.52 & 13.77 & $43900 \pm 1000$ & $6.07 \pm 0.17$ \\
\object{MSP\,188} & 10:24:01.2 & $-57$:45:31 & O7:V & O4\,V-III   & O4\,V-III      & 13.41 & 1.37 & 5.12 & 14.10 & $42600 \pm 1000$ & $5.93 \pm 0.16$ \\
\object{MSP\,199} & 10:24:02.7 & $-57$:45:34 &    & O3\,V       & O3-4\,V        & 14.37 & 1.40 & 5.21 & 14.88 & $43900 \pm 1000$ & $5.61 \pm 0.17$ \\
\object{MSP\,203} & 10:24:02.3 & $-57$:45:35 & O7:V & O6\,V-III   & O6\,V-III      & 13.31 & 1.37 & 5.10 & 13.59 & $38600 \pm 800$  & $5.84 \pm 0.16$ \\
\object{MSP\,263} & 10:24:01.5 & $-57$:45:57 &     & O6.5\,V     & O6\,V          & 15.01 & 1.61 & 5.84 & 14.09 & $38350 \pm 500$  & $5.45 \pm 0.16$ \\
\hline
\end{tabular}
\end{table*}

\subsection{Early-type stars \label{specttypeO}}
Twelve of our targets display spectra typical of O-type stars (see Fig.\,\ref{Otype}). The spectral types were derived from the ratio of the equivalent widths of the He\,{\sc i} $\lambda$\,4471 and He\,{\sc ii} $\lambda$\,4542 classification lines by means of the criteria first introduced by Conti (\cite{Conti}, see also Mathys \cite{Mathys}). The results are displayed in Table\,\ref{types}. For star MSP\,203, we were only able to derive an upper limit on the equivalent width of the He\,{\sc i} $\lambda$\,4471 line.
All stars have rather strong He\,{\sc ii} $\lambda$\,4686 absorption lines (equivalent widths in the range 0.4 to 1.1\,\AA, the extremes corresponding to MSP\,167 and 263 respectively) suggesting they are main-sequence or giant objects. Mathys (\cite{Mathys}) proposed $EW = 0.56$\,\AA\ as the demarcation between weak and strong He\,{\sc ii} $\lambda$\,4686 absorption corresponding to luminosity classes III and V respectively. Using this criterion, we have assigned the luminosity classes in Table\,\ref{types}.

We then compared our spectra to the digital atlas of Walborn \& Fitzpatrick (\cite{WF}) and Walborn et al.\ (\cite{O2}). The results are also indicated in Table\,\ref{types}. The classifications according to the two criteria agree to better than one subtype. We note that because the He\,{\sc i} $\lambda$\,4471 classification line is quite weak or broad in the spectra of MSP\,167, 175 and 203, these objects could also be classified as O5 rather than O6 stars. Henceforth, we adopt the conservative hypothesis that these objects are actually O6 stars, but we will nevertheless consider the implications of a somewhat earlier classification for these stars when discussing the ionization of RCW\,49.

Three stars (MSP\,18, 182 and 183) display a strong He\,{\sc ii} $\lambda$\,4686 absorption accompanied by weak N\,{\sc iii} $\lambda\lambda$\,4634-40 emission, which leads to an ((f)) classification. The earliest star in our sample is MSP\,183 which displays prominent N\,{\sc v} $\lambda\lambda$\,4604, 4620 absorption lines. Except for the weaker N\,{\sc iii} $\lambda\lambda$\,4634-4640 lines, the spectrum of MSP\,183 very much resembles that of HD\,64568 (O3\,V((f$^*$)), Walborn et al.\ \cite{O2}). We therefore classify MSP\,183 as O3\,V((f)). 
\begin{figure}[h]
%\resizebox{8.7cm}{!}{\includegraphics{6495fig3.ps}}
\caption{Medium resolution spectra of O stars in Westerlund\,2. Note the very strong diffuse interstellar band (DIB) at 4430\AA. Some residual nebular H$\beta$ line emission affects the spectra of MSP\,151, 182 and 199. \label{Otype}}
\end{figure}

On average the spectral types that we determine are significantly earlier than those inferred by Moffat et al.\ (\cite{MSP}). The implications of these revised classifications are discussed in Sect.\,\ref{link}. Uzpen et al.\ (\cite{Uzpen}) recently inferred an O4\,V((f)) spectral type for MSP\,18. Our data have a superior resolution (1\,\AA) than their observations (4.2\,\AA) and are thus better suited to measure the strength of the weak He\,{\sc i} classification lines. We therefore favor the O5\,V((f)) spectral classification for this star.

Though our data set is obviously not ideal for this purpose, we note that our spectra do not reveal strong evidence for multiplicity. None of the stars has an obvious composite spectrum nor displays assymmetric, very broad or double lines that could make it a spectroscopic binary candidate. A higher resolution, high S/N spectroscopic monitoring of these stars is however needed to confirm or infirm this impression.  

In Table\,\ref{types}, we list the observed $V$ magnitudes and the $B-V$ colours for the spectroscopically identified O-type stars in our sample. For stars that are well isolated on the sky, the aperture and PSF photometry agree very well with each other and we adopted the magnitude obtained with an aperture of 8 pixels (i.e.\ 3.0\,\arcsec) radius which has the smallest errors. On the other hand, for stars in the crowded core of the cluster, it was mandatory to use the PSF photometry instead. Comparing the observed colour indices with the intrinsic colours of Martins \& Plez (\cite{MP}), we have inferred the extinction $A_V$, adopting $R_V = 3.1$. Assuming that these objects have absolute magnitudes typical of main-sequence or giant O-type stars (Martins \& Plez \cite{MP}), we estimate an average distance modulus of $14.47 \pm 0.65$ (1-$\sigma$ standard deviation). If we exclude the three stars with the lowest (MSP\,18, 183 and 203) and the star with the largest (MSP\,167) value of the distance modulus, we get on average $DM = 14.60 \pm 0.43$ in excellent agreement with the value ($14.5 \pm 0.3$) of Moffat et al.\ (\cite{MSP}) based on their $UBV$ photometry and ZAMS fitting. This suggests therefore a distance of $(8.3 \pm 1.6)$\,kpc towards Westerlund\,2 in very good agreement with the distance estimate of $\sim 8.0$\,kpc towards WR\,20a (see Rauw et al.\ \cite{Rauw05} and Sect.\,\ref{20a}). We note that adopting the intrinsic colours from Schmidt-Kaler (\cite{SK}) and the absolute magnitudes from Martins et al.\ (\cite{Martins}) would reduce the distance modulus by 0.23\,mag (i.e.\ a reduction of the distance by 800\,pc).

\section{Photometric variability}
\subsection{WR\,20a \label{20a}}
\begin{figure}[htb]
\resizebox{8.5cm}{!}{\includegraphics{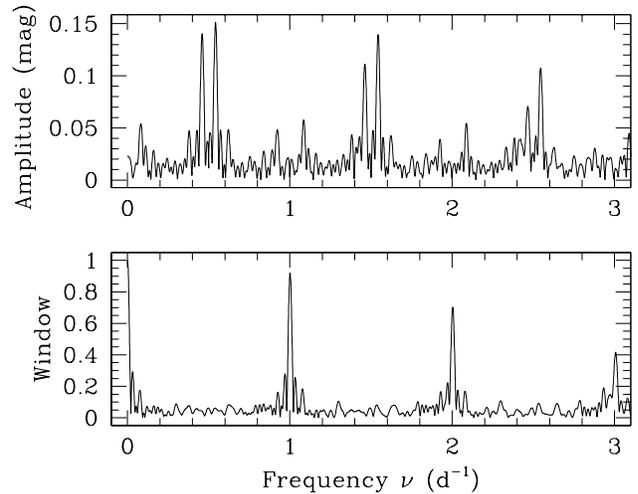}}
\caption{Periodogram showing the half amplitude of the best-fit sine wave (top panel) and spectral window (bottom panel) derived from the Fourier analysis of the photometric time series of WR\,20a. The results are shown for the measurements in the $B$ filter (8 pixels radius aperture photometry). For the $V$ filter, the results are identical. The highest peak in the periodogram is located at 0.5426\,d$^{-1}$.\label{fourier20a}}
\end{figure}

As pointed out in the introduction, one of the initial motivations of our observing campaign was the study of the photometric eclipses of the extremely massive binary WR\,20a. Bonanos et al.\ (\cite{bonanos}) reported $I$ band observations showing that the system displays indeed eclipses on the orbital period of 3.686\,days. The Fourier periodogram (computed using the method of Heck et al.\ \cite{HMM}, revised by Gosset et al.\ \cite{wr30a}) of our photometric time series of this system displays the highest peak at $\nu'_1 = 0.5426$\,d$^{-1}$ for both filters ($B$ and $V$) and irrespective of whether we consider the aperture or PSF photometry (see e.g.\ Fig.\,\ref{fourier20a}). The natural width of the peaks in the Fourier periodogram of our time series is equal to 0.0222\,d$^{-1}$. Assuming that the position of the highest peak can be determined with an accuracy of 5\% of this width, we infer an uncertainty of 0.0011\,d$^{-1}$ on $\nu'_1$.

Since the two components of WR\,20a have roughly equal temperature and brightness, the eclipses are of essentially equal depth and shape. Thus $\nu'_1$ actually corresponds to twice the orbital frequency. Therefore, our data yield an orbital period of $(3.6859 \pm 0.0075)$\,days, in very good agreement with the value ($3.686 \pm 0.010$\,days) found by Bonanos et al.\ (\cite{bonanos}). The whole periodogram (Fig.\,\ref{fourier20a}) can be interpreted on the basis of this sole periodicity. Indeed, no peak is visible at $\nu \sim 0.27$\,d$^{-1}$ because most of the power is contained in the $\nu'_1$ harmonic. The higher harmonic at $\nu \sim 1.08$\,d$^{-1}$ is induced by the shape of the eclipses whereas the harmonics at $\nu \sim 0.82$\,d$^{-1}$ (not to be confused with $\nu = 0.93$\,d$^{-1}$, the one-day alias of $1.08$\,d$^{-1}$) is related to the slight difference in eclipse depths.

Folding the light curve of WR\,20a with the derived period, we have evaluated the magnitudes of the system at phases outside of the eclipses: $B = 14.999 \pm 0.021$, $V = 13.416 \pm 0.024$ and $I = 10.658 \pm 0.015$ (the latter value was computed from the Bonanos et al.\ data). Our estimate of the period can be further refined by comparing our light curves with that of Bonanos et al. Subtracting the above outside-eclipse magnitudes from the observed light curves, we can combine all the data and perform a Fourier analysis. In this way, we find that the highest peak in the periodogram has $\nu_1 = (0.54277 \pm 0.00019)$\,d$^{-1}$. Therefore, the newly determined value of the orbital period is $(3.68475 \pm 0.00129)$\,days. The value of the uncertainty is probably too conservative: comparing the time of primary minimum determined by Bonanos et al.\ (\cite{bonanos}) with our best estimate of the primary minimum results in a more realistic estimate of the uncertainty on the orbital period of $0.00020$\,days. 

The light curves of WR\,20a folded with the above period and adopting a time of primary minimum of HJD\,2453330.930 are displayed in Fig.\,\ref{lc20a}. We have fitted our $B$ and $V$ light curve along with the $I$ band light curve from Bonanos et al.\ (\cite{bonanos}) using the {\sc nightfall}\footnote{ http://www.hs.uni-hamburg.de/DE/Ins/per/Wichmann/\\Nightfall.html\\ Note that the {\sc nightfall} code does not include stellar wind effects, but given the proximity of the stars in WR\,20a, the light curve should be dominated by the geometrical effects.} software developed and maintained by R.\ Wichmann, M.\ Kuster \& P.\ Risse.
We fixed the mass ratio at the value ($m_{\rm sec}/m_{\rm prim} = 0.99$) derived from the radial velocity curve. There exist a number of combinations of the model parameters that fit the light curve but do not provide a physically self-consistent description of the system. For instance, if the filling factors (defined as the ratio of the polar radius of the star to the polar radius of the Roche lobe) of both components are allowed to differ in the fit of the light curve, the luminosities of the primary and secondary star are found to differ by more than 50\% which is clearly at odds with the fact that the two stars have identical spectral types and spectral features of identical strength (Rauw et al.\ \cite{Rauw04}, \cite{Rauw05}). We have therefore constrained some of the fitting parameters. For instance, we searched for models which have primary and secondary temperatures in agreement with the value ($43000 \pm 2000$\,K) derived from an analysis with a model atmosphere code (see Rauw et al.\ \cite{Rauw05}). Following the same strategy as Bonanos et al.\ (\cite{bonanos}), we have assumed that both stars have identical (Roche lobe) filling factors.  For a grid of values of the filling factors and the primary (resp.\ secondary) temperature, we iterated the {\sc nightfall} code for the best fit inclination and secondary (resp.\ primary) temperature. The best constrained model parameter is the orbital inclination which quickly converges to $i = (74.5 \pm 1.0)^{\circ}$ in excellent agreement with the value derived by Bonanos et al.\ (\cite{bonanos}). Moreover, the shape of the wings of the eclipses and the light variations outside of the eclipses require that the filling factor be between $0.89$ and $0.92$ with a preferred value of $0.91$. The absolute temperatures of the components cannot be constrained by the analysis of the light curve. In principle, the depths of the eclipses constrain the relative surface brightnesses and hence the relative temperatures. Under the above assumptions of equal radii, we find that the primary should be hotter than the secondary by about 2000 -- 3000\,K (assuming a primary temperature in the range found from the spectroscopic analysis), although models with identical temperatures for both components are of almost equivalent quality. The best fit light curves for a filling factor of $0.91$ for both stars, an orbital inclination of $74.5^{\circ}$, and temperatures of T$_{\rm eff, p} = 43000$\,K and T$_{\rm eff, s} = 40500$\,K are illustrated in Fig.\,\ref{lc20a}.

\begin{figure}[htb]
%\resizebox{8.5cm}{!}{\includegraphics{6495fig5.ps}}
\caption{Light curve of WR\,20a in the $B$, $V$ and $I$ (the latter data from Bonanos et al.\ \cite{bonanos}) filters folded with the 3.68475\,day period and adopting a time of primary minimum of HJD\,2453330.930. The $B$ and $V$ data correspond to the photometry obtained through PSF fitting and are given in Table 2 (available in electronic form at the CDS). Filled symbols stand for data with a {\tt FWHM} of the PSF of less than 4.0 pixels, whilst the open symbols yield data taken under poorer seeing conditions. The continuous curve represents the best fit theoretical light curve determined with the {\sc nightfall} code for a filling factor of $0.91$ for both stars, an orbital inclination of $74.5^{\circ}$, and temperatures of T$_{\rm eff, p} = 43000$\,K and T$_{\rm eff, s} = 40500$\,K.\label{lc20a}}
\end{figure}

The parameters of the above solution correspond to an average stellar radius of $(18.7 \pm 0.9)$\,R$_{\odot}$, roughly 5\% smaller than the values found by Bonanos et al.\ (\cite{bonanos}). If the primary temperature amounts to $(43000 \pm 2000)$\,K as found by Rauw et al.\ (\cite{Rauw05}), then its bolometric luminosity would be $\log{L/L_{\odot}} = 6.03 \pm 0.09$, whilst for the secondary we obtain $\log{L/L_{\odot}} = 5.93 \pm 0.10$. For the entire binary system, we thus find $\log{L/L_{\odot}} = 6.28 \pm 0.10$. With the bolometric correction of $-3.91$ (Rauw et al.\ \cite{Rauw05}), we therefore infer an absolute $V$ magnitude of $M_V = -7.04 \pm 0.25$. The out of eclipse magnitudes listed above yield $B - V = 1.58 \pm 0.03$, hence $E(B - V) = 1.91 \pm 0.03$ (evaluated using the intrinsic colour index from Rauw et al.\ \cite{Rauw05}) and $A_V = 5.93 \pm 0.09$. This leads to $DM = 14.52 \pm 0.27$ corresponding to $d = (8.0 \pm 1.0)$\,kpc. This result concurs with the estimate of Rauw et al.\ (\cite{Rauw05}; $d = (7.9 \pm 0.6)$\,kpc), and agrees perfectly well with the estimate of the distance of the O-type star population presented above. We will come back to this issue in Sect.\,\ref{link}.

\begin{figure*}[t!]
\begin{minipage}{5.8cm}
\resizebox{6.0cm}{!}{\includegraphics{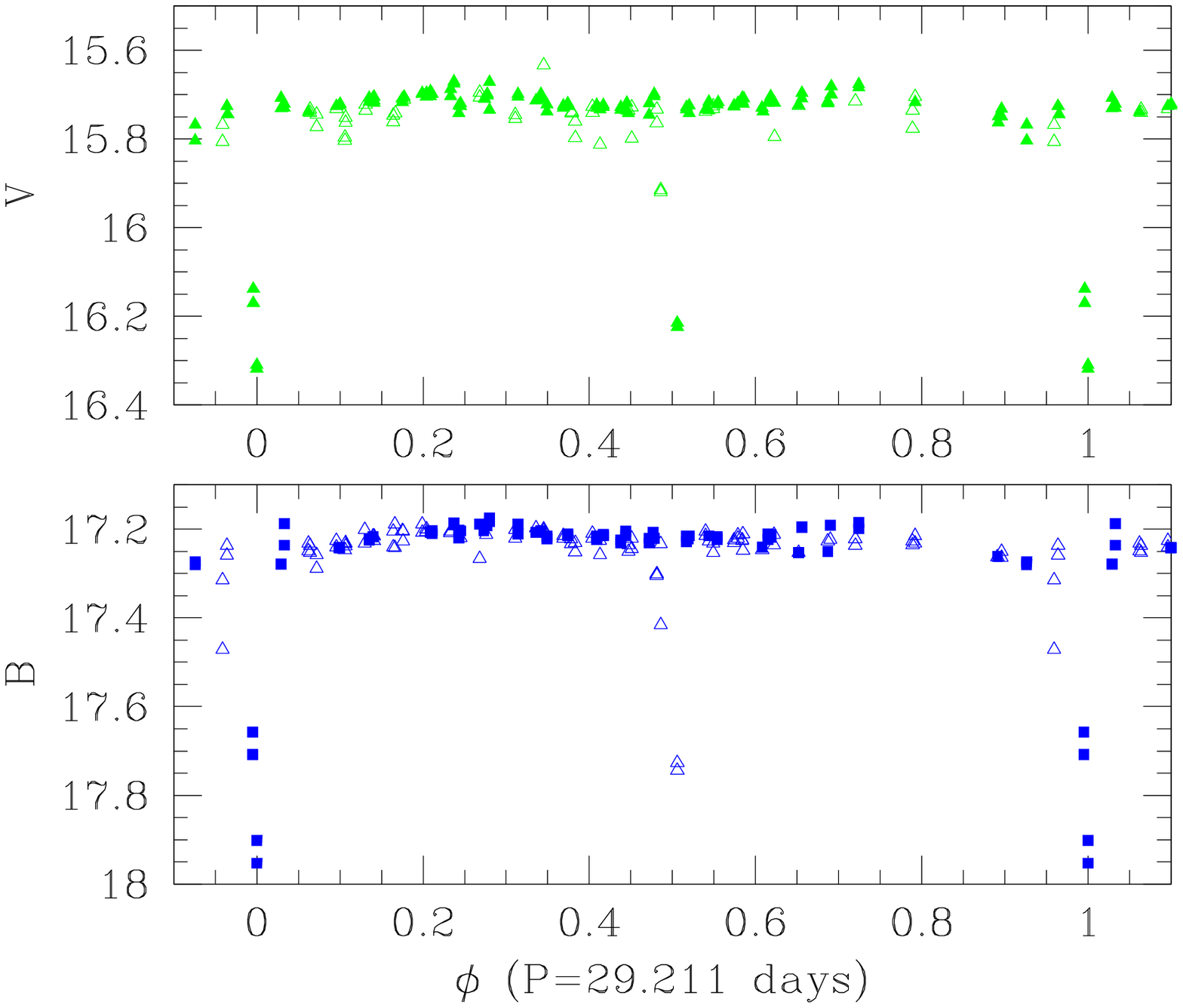}}
\end{minipage}
\hfill
\begin{minipage}{5.8cm}
\resizebox{6.0cm}{!}{\includegraphics{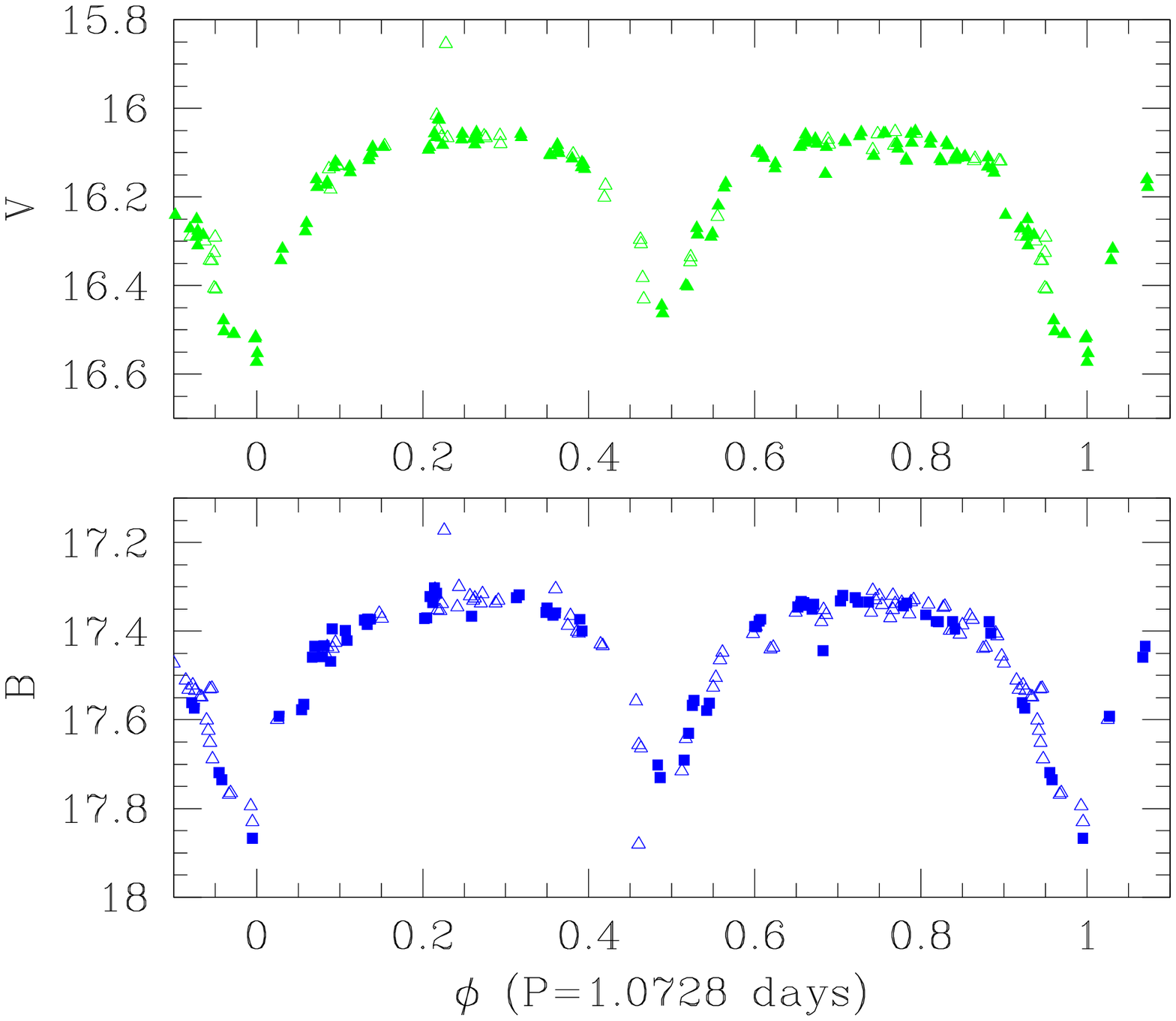}}
\end{minipage}
\hfill
\begin{minipage}{5.8cm}
\resizebox{6.0cm}{!}{\includegraphics{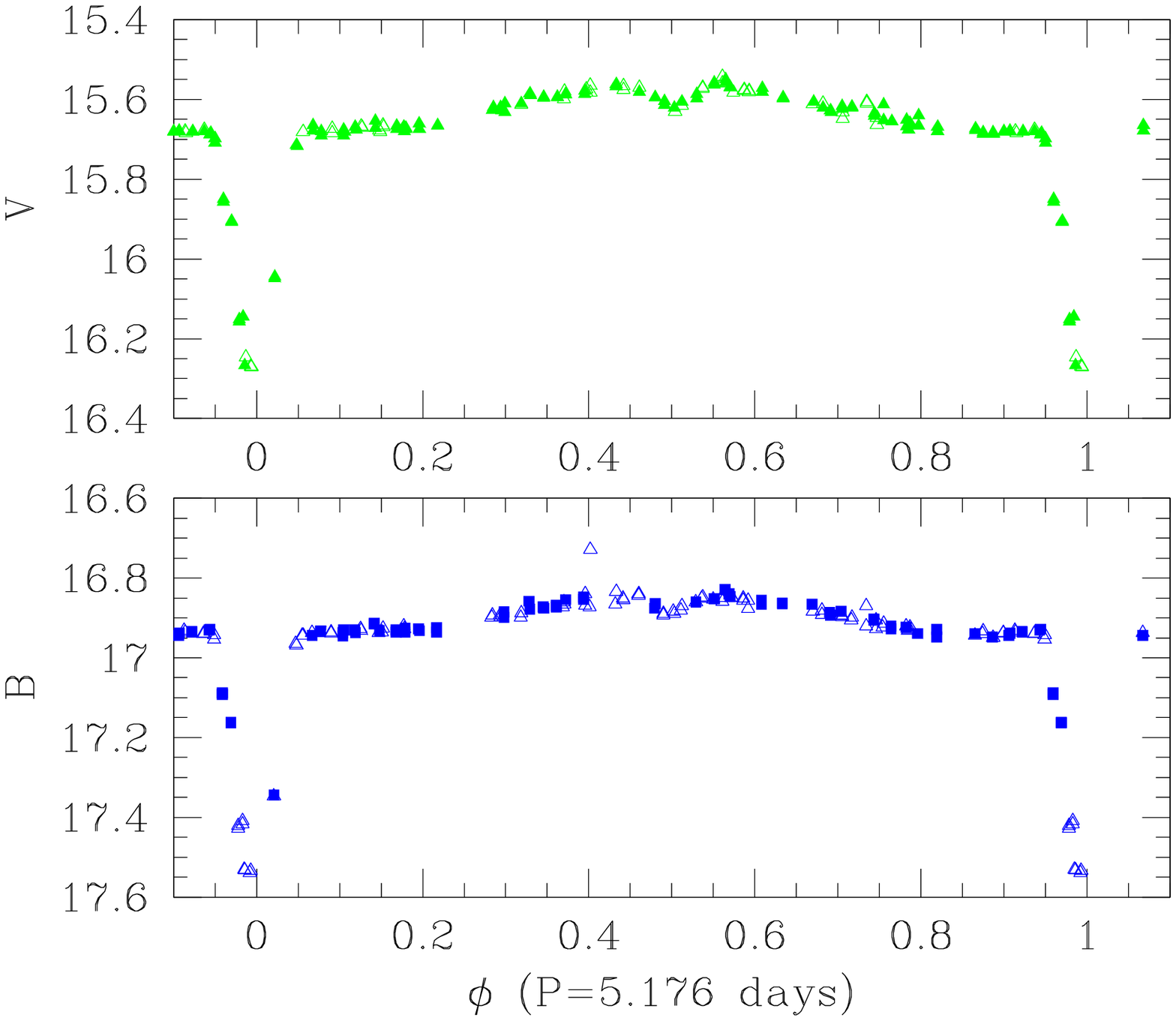}}
\end{minipage}
\caption{{\it Left:} light curves of MSP\,223 as obtained from our $B$ and $V$ PSF photometry time series folded with a period of 29.211\,days and $T_0 = 2453369.85$. The filled and open symbols stand for data obtained with a seeing of less than and more than 4.0 pixels ({\sc fwhm}) respectively. {\it Middle:} same as left panel, but for MSP\,96, a period of 1.0728\,days and $T_0 = 2453370.6973$. {\it Right:} same as left panel, but for MSP\,44, a period of 5.176\,days and $T_0 = 2453371.75$.
\label{otherlc}}
\end{figure*}
Although the periodogram of Fig.\,\ref{fourier20a} can be fully explained by the orbital period and its harmonics, we decided to search for additional variations, in order to extract the maximum information from the data. Moreover, it is reasonable to think that such massive stars might display additional variations such as pulsations (see Sterken \& Breysacher \cite{Sterken}, and Fig.\,3 of Pamyatnykh \cite{Pamyatnykh}). Consequently, we have whitened the $B$, $V$ photometry time series for the binary variations by removing the orbital period and its harmonics. The time series of the residuals have again been Fourier analysed. No peak exceeding a semi-amplitude of 0.008\,mag has been found, indicating that no other periodicity is present in the photometric variations of WR\,20a at this level. 
\subsection{O-type stars}
Using our photometric time series, we have searched for variability among the objects identified as early-type O-stars in our spectroscopic campaign. For targets that are rather well isolated, the aperture photometry yields 1-$\sigma$ dispersions on the $B$ and $V$ magnitudes of order 0.005 to 0.010\,mag. A Fourier analysis of our time series, excluding the data points with the poorest seeing reveals no significant peak in the periodogram. The highest peak is usually found near $\nu = 1$\,d$^{-1}$ which is actually due to the interplay between the sampling of our time series and a few deviating data points.

For objects in the crowded region of the cluster core, we find that the variations of the aperture photometry time series reflect the variations of the seeing during the observing campaign as could be expected. Excluding those points with the poorest observing conditions from the aperture photometry time series, we find that there is no photometric variability exceeding the 1-$\sigma$ level of 0.010\,mag. The Fourier analysis yields the same conclusions as for the isolated objects. Although the PSF photometry provides more accurate absolute photometry, PSF photometry time series usually yield somewhat larger 1-$\sigma$ dispersions. 
We conclude that none of the early O-type stars in the cluster core displays a significant variability and in particular, we find no evidence for photometric eclipses. 

\subsection{Other variables in the field of view \label{others}}
We have searched our photometric time series for additional variables. The light curves of those objects that display a photometric variability exceeding significantly the level of photon noise in the $B$ and $V$ band were extracted and were analysed with the Fourier technique over the frequency range between 0.0 and 5.0\,d$^{-1}$. Only data points obtained when the seeing was better than 4 pixels (i.e.\ 1.5\,\arcsec) were included in this process. For most objects, the Fourier analysis yields the highest peaks at frequencies near 0 or an integer number of day$^{-1}$. These features stem from the combination of the actual sampling of the time series with a few deviating data points and do not reflect a genuine periodic behaviour. The only exceptions are the three stars discussed below.

\object{MSP\,223} (10:23:59.2 $-57$:45:40, J2000) is an apparently red star ($V = 15.75$, $B - V = 1.50$) that shows drops in its light curve that are consistently seen in the $B$ and $V$ filter and occur on a recurrence time of 14.61\,days. The variations are reminiscent of an eclipsing binary. In this case, the system either lacks a secondary eclipse (maybe as a result of an eccentric orbit or of an insufficient sampling of the light curve near that phase) or the true orbital period could be 29.21\,days. We folded the light curve with this latter period and adopting a time of primary minimum of HJD\,2453369.85. The eclipses are rather narrow and the secondary eclipse could be slightly shallower than the primary one. The system presents some slight slow variations outside the eclipses (see Fig.\,\ref{otherlc}). Though the photometric data are not sufficient to provide a full solution, test calculations with the {\sc nightfall} code suggest that the system consists of a pair of stars of comparable surface brightness that are well inside their Roche lobe and are seen under a high inclination angle. The variations outside eclipse cannot be explained by ellipsoidal variations and rather suggest the existence of a non-uniform surface brightness for at least one component. MSP\,223 could thus be a kind of magnetically active binary.

The periodogram of the photometry of \object{MSP\,96} (10:24:01.0 $-57$:45:06, J2000; $V = 16.17$, $B - V = 1.27$) displays a prominent peak at 1.864\,day$^{-1}$, although its one-day aliases are only slightly weaker. Again the light curve suggests an eclipsing binary and the true period is probably 1.0728\,day. The eclipses are rather broad (see Fig.\,\ref{otherlc}) and the ellipsoidal variations suggest that at least one of the two stars must be close to filling its Roche lobe. Also, the different depths of the eclipses indicate that one of the stars must be significantly hotter than its companion. 

\begin{figure}[htb]
\resizebox{8.5cm}{!}{\includegraphics{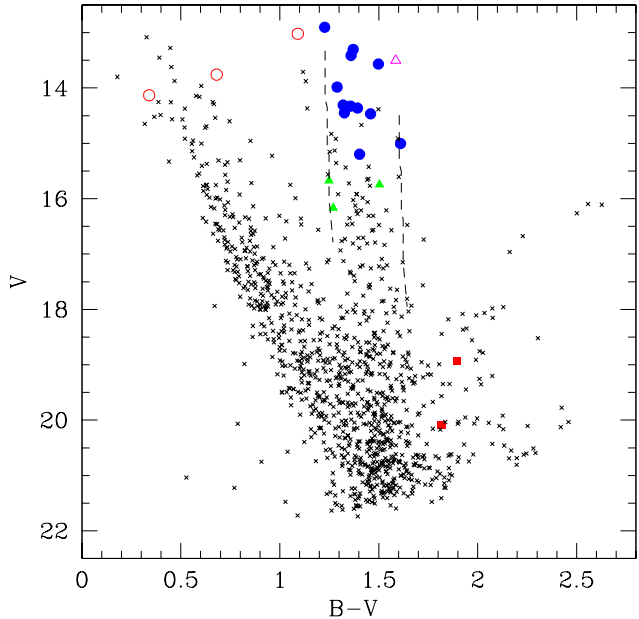}}
\caption{Colour-magnitude diagram of the stars in the ANDICAM field of view around Westerlund\,2. The filled circles and the open triangle correspond to the twelve O-type stars observed in spectroscopy (see Table \ref{types}) and the WR\,20a binary, respectively. The dashed lines correspond to the main-sequence between spectral types O3 and B2 for a distance modulus of 14.52 and reddened with $A_V = 4.68$ (leftmost curve) or $5.84$ (rightmost curve) and assuming $R_V = 3.1$. The open circles yield the late-type foreground objects MSP\,91, 158 and 218 whilst the filled triangles indicate the three eclipsing binaries discussed in Sect.\,\ref{others}. Finally, the two squares correspond to the two non-periodic variables.\label{coulmag}}
\end{figure}
The periodograms of \object{MSP\,44} (10:24:00.5 $-57$:44:45, J2000; $V = 15.67$, $B - V = 1.25$) display their highest peak at 0.1932\,day$^{-1}$, although the $1-\nu$ alias is almost equally strong. The light curve folded with a period of 5.176\,days and $T_0 = 2453371.75$ displays a strong minimum at phase 0.0 as well as a much weaker secondary minimum at phase 0.5. Outside these rather narrow minima the light curve is not flat, but displays a strong modulation (see Fig.\,\ref{otherlc}). As for MSP\,223, the narrow eclipses are the signature of a pair of relatively compact stars seen under high inclination, whilst the strong modulation of the flux outside eclipse requires a non-uniform surface brightness (spots or reflection effects) probably for both components. The light curve of MSP\,44 is reminiscent of those of RS\,CVn systems therefore making this a candidate for a magnetically active binary system. Alternatively, MSP\,44 could be a detached binary with components that differ quite substantially in temperature. The seemingly symmetric variability outside of the eclipse would then result from a strong reflection effect or surface heating.

To further constrain the nature of MSP\,44, 96 and 223, we examine the $(B-V,V)$ colour-magnitude diagram of the stars\footnote{Tables 3 and 4 with the light curves of the eclipsing binaries and the mean $(B-V,V)$ data are provided at the CDS.} in the ANDICAM field of view around Westerlund\,2 (see Fig.\,\ref{coulmag}). Figure\,\ref{coulmag} clearly reveals two populations of stars: the foreground field stars which have relatively blue colours and the cluster stars which have an $E(B-V)$ colour excess between 1.51 and 1.88. The O-type stars and WR\,20a fall in the region of  the cluster main-sequence (see also Sect.\,\ref{hrd}). Two of the three foreground stars MSP\,91 and 218 clearly occupy the locus of field stars. The third one (MSP\,158, G8\,III) falls close to the cluster main sequence, suggesting that the locus of the cluster stars might be polluted by cluster non-members. The three eclipsing binaries fall on the locus of the cluster main-sequence. Moffat et al.\ (\cite{MSP}) actually labelled all three stars as probable OB members of Westerlund\,2 based on their $UBV$ photometry. In this case, the primary components of these binaries would be late O or early B-type stars which would contradict our suggestion of two of these systems being magnetically active late-type binaries. A spectroscopic follow up of these objects will not only help clarify their nature, but will also bring valuable constraints on the general cluster properties.

Finally, Kolmogorov-Smirnov and $\chi^2$ tests were also performed to identify non-periodic variables. Two more stars emerged from these tests and both were found to display a strong apparently non-periodic variability. These objects are located at 10:23:58 $-57$:47.7 and 10:24:00 $-57$:44.9 (J2000) and have magnitudes and colours of $V = 20.01$, $B - V = 1.83$ and $V = 18.93$, $B - V = 1.88$, respectively. Both objects are about 1\,mag fainter (in $B$ and $V$) at the beginning of our campaign than later and these `minima' last about 14 and 18 days respectively. No other event of this type is seen over the rest of the campaign. These objects are either irregular or long-term (period $> 45$\,days) variables or eclipsing binaries with periods very close to 1 or 2\,days.  
\section{The link between Westerlund\,2, WR\,20a and RCW\,49 \label{link}}
\subsection{The distance \label{distance}}
The distance towards Westerlund\,2 and RCW\,49 has been a very controversial issue (see e.g.\ the discussions in Churchwell et al.\ \cite{Churchwell04} and Rauw et al.\ \cite{Rauw05}) and distance estimates for the cluster range from about 2.5\,kpc to 8\,kpc. The line of sight towards RCW\,49 is roughly tangential to the Carina arm, leading to a wide range of possible distances. Using a galactic rotation curve along with H\,{\sc i} line velocity measurements, Russeil (\cite{Russeil}) recently determined a kinematic distance of $4.7^{+0.6}_{-0.2}$\,kpc towards RCW\,49. Russeil cautioned however that velocity deviations from the circular Galactic rotation model exist all along the Carina spiral arm. 

Based on spectro-photometric parallaxes of the early O-type stars, we have estimated $d = (8.3 \pm 1.6)$\,kpc, whilst the analysis of the light curve of the eclipsing binary WR\,20a yields $d = (8.0 \pm 1.0)$\,kpc. The weighted mean of these two distance estimates amounts to $d = (8.0 \pm 1.4)$\,kpc. 
These results clearly rule out a distance below $\sim 6$\,kpc. Moreover, the excellent agreement between our two distance determinations indicates that WR\,20a very likely belongs to Westerlund\,2. Various effects that could bias our distance determinations are briefly discussed in Appendix \ref{append}.

Based on {\it Spitzer} data, Uzpen et al.\ (\cite{Uzpen}) found that MSP\,18 displays an IR excess which they attributed to thermal bremsstrahlung from the stellar wind. These authors fitted the $V$, {\it Spitzer} and 2MASS photometry with a Kurucz ATLAS9 model. In this procedure, the scaling factor of the surface flux depends on the angular diameter (i.e.\ on the ratio of the stellar radius and the distance) of the star. In this way, they inferred a spectro-photometric distance of $3233^{+540}_{-535}$\,pc and an extinction of $A_V = 5.63^{+0.01}_{-0.30}$. These authors suggest therefore that the actual distance of Westerlund\,2 is 3.2\,kpc. The extinction is significantly larger than our value (4.68), whilst their distance is less than 50\% of our best estimate of the cluster distance. 
To understand the origin of this discrepancy, we fitted our $B$, $V$ magnitudes along with the $U$ magnitude from Moffat et al.\ (\cite{MSP}) and the 2MASS photometry using a simple (static, plane-parallel atmosphere in LTE) model atmosphere code\footnote{The software, developed by S.\ Jeffery at Armagh Observatory, is available at http://star.arm.ac.uk/$\sim$csj/software.store.}. Test fits including the 2MASS data suggest that the large $A_V$ value of Uzpen et al.\ (\cite{Uzpen}) is likely due to the near-IR domain already being affected by the IR excess emission. Freezing the reddening at the value of $E(B - V) = 1.56$ found above, the temperature at 40\,kK compatible with the optical spectrum and fitting only the $U\,B\,V$ data clearly confirms this idea. The angular diameter derived from the latter fit amounts to $6 \times 10^{-11}$\,rad. Assuming a typical radius of 12.0\,R$_{\odot}$ for the O5\,V-III star then yields a distance of 9.0\,kpc. We thus conclude that the discrepancy between the Uzpen et al.\ (\cite{Uzpen}) result and our estimate stems from the near-IR photometry which is already affected by excess emission.

In summary, we find that the distance of WR\,20a and Westerlund\,2 is probably 
$(8.0 \pm 1.4)$\,kpc. 

\subsection{The Hertzsprung-Russell diagram of Westerlund\,2 \label{hrd}}
To position the O-type stars in the Hertzsprung-Russell diagram, we have adopted the `observed' spectral-type -- effective temperature calibration and bolometric corrections of Martins et al.\ (\cite{Martins}) and Martins \& Plez (\cite{MP}), respectively. The observed $V$ magnitude and $A_V$ extinction of each star were taken from Table \ref{types}. The bolometric luminosities are quoted in the last column of Table \ref{types} and were evaluated accounting for the uncertainties in spectral type and luminosity class and assuming a distance of $(8.0 \pm 1.4)$\,kpc. 

\begin{figure}[htb]
\resizebox{8.5cm}{!}{\includegraphics{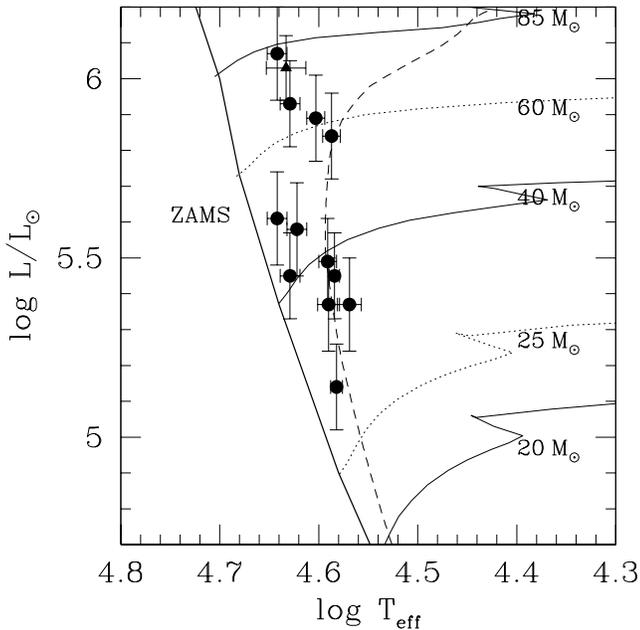}}
\caption{Hertzsprung-Russell diagram of the O-type stars in Westerlund\,2.
Evolutionary tracks are from Schaller et al.\ (\cite{Schaller}). The dashed line yields the 2.5\,Myr isochrone corresponding to an upper limit on the age of the O-stars. The filled triangle indicates the position of the WN6ha stars of WR\,20a.\label{hrdwd2}}
\end{figure}

In Fig.\,\ref{hrdwd2}, we compare the location of the O-stars in the Hertzsprung-Russell diagram with the evolutionary tracks and isochrones from Schaller et al.\ (\cite{Schaller}) for solar metallicity. Keeping in mind the possibility that some of the luminosities might be overestimated due to binarity, we note that the O-stars that we have observed should have masses ranging from about 30 to 85\,M$_{\odot}$ and ages of less than $\sim 2.5$\,Myr. 

\subsection{The ionization of RCW\,49}
Several authors have presented radio observations of RCW\,49. The continuum radio emission of H\,{\sc ii} regions is free-free emission that can, in principle, be used to evaluate the minimum number of Lyman continuum photons. In the case of RCW\,49, the observed radio luminosity (evaluated for a distance near 8\,kpc) was usually found to be too large compared to the Lyman ionizing flux from the known O-star population. For instance, Whiteoak \& Uchida (\cite{WU}) evaluated a flux density of 210\,Jy at 0.843\,GHz for the entire RCW\,49 complex, whilst the flux density of the shell around Westerlund\,2 was found equal to 110\,Jy. Based on these numbers and the census of O-type stars of Moffat et al.\ (\cite{MSP}), Whiteoak \& Uchida (\cite{WU}) further argued that the Lyman continuum flux of the early-type stars could only explain the observed radio emission of RCW\,49 if the distance were of order 2.3\,kpc.
 
In this context, we emphasize that adopting a lower distance for the nebula and the cluster does not solve the issue of the missing ionizing photons. In fact, both, the radio flux and hence the required number of ionizing photons on the one hand and the bolometric luminosity of the stars on the other hand scale with the distance squared. Assuming a lower distance as was done by Whiteoak \& Uchida (\cite{WU}) would thus imply a drastic under-luminosity of the stars compared to typical objects of the same spectral type. Therefore, the ionizing flux would also be reduced compared to the values for typical stars. 

For a distance of 8.0\,kpc, the number of Lyman continuum photons required to account for the total 0.843\,GHz flux of RCW\,49 and the flux of the shell near Westerlund\,2 as quoted by Whiteoak \& Uchida (\cite{WU}) would be $11.5 \times 10^{50}$ and $6.0 \times 10^{50}$\,s$^{-1}$, respectively. In a similar way, scaling the result of Churchwell et al.\ (\cite{Churchwell04}) to our preferred distance estimate yields a minimum UV photon luminosity of $20.7 \times 10^{50}$\,s$^{-1}$ to account for the 5\,GHz flux density of 335\,Jy (as determined by Goss \& Shaver \cite{GS}). We note that lower determinations of the 5\,GHz flux can be found in the literature: 178.8\,Jy (Smith et al.\ \cite{SBM}) and 263.2\,Jy (Kuchar \& Clark \cite{KC}), which imply photon numbers in the range $11.1$ -- $16.3 \times 10^{50}$\,s$^{-1}$. On average, we find $(14.9 \pm 4.5)\times 10^{50}$\,s$^{-1}$. These numbers actually represent a lower limit on the ionizing flux since they neglect the possible leakage of some UV radiation from the nebula (see Churchwell et al.\ \cite{Churchwell04}) and they were estimated for a dust-free nebula using the formalism of Mezger et al.\ (\cite{Mezger}). Near- and mid-IR images of the RCW\,49 complex reveal however the existence of dust in this nebula, although the stellar winds and radiations of the O-type stars have cleared the region immediately surrounding Westerlund\,2 (e.g.\ Conti \& Crowther \cite{CC} and Churchwell et al.\ \cite{Churchwell04}). Using the formalism of Smith et al.\ (\cite{SBM}) to account for the impact of dust, we estimate that about 58\% of the UV continuum flux is actually used to ionize the nebula. As a result, the total number of Lyman photons required to maintain the ionization of RCW\,49 would have to be $(25.7 \pm 7.9)\times 10^{50}$\,s$^{-1}$.\\

With the newly determined spectral types and bolometric luminosities at hand, we have evaluated the number of ionizing photons produced by the O-type stars of Westerlund\,2. To this aim, we have scaled the numbers of Lyman ionizing photons emitted per second ($Q_0$) as tabulated by Martins et al.\ (\cite{Martins}) by the ratio of the above determined bolometric luminosity divided by the 'typical' theoretical bolometric luminosity. For a distance of 8.0\,kpc, the total number of Lyman ionizing photons produced by the twelve O-type stars in our sample amounts to $3.2 \times 10^{50}$\,s$^{-1}$. Assuming that stars in the spectral range O6.5 -- O3 have masses between 28 and 60\,M$_{\odot}$, and adopting a Salpeter IMF, we have estimated that Westerlund\,2 should harbour about 20 stars with spectral types in the range O9.5 -- O7 (masses of 15.5 -- 28\,M$_{\odot}$). These late O-type stars would contribute about $7 \times 10^{49}$\,s$^{-1}$ (i.e.\ $\sim 20$\% of the early-type O-stars) to the total number of Lyman ionizing photons. Adding the ionizing flux of the WR\,20a pair of WN6ha stars ($1.45 \times 10^{50}$\,s$^{-1}$, Rauw et al.\ \cite{Rauw05}) we arrive at a total of $5.4 \times 10^{50}$\,s$^{-1}$, i.e.\ a fifth of the photon flux required to explain the radio emission of RCW\,49 (or slightly more than one third of the required flux if we neglect the impact of the dust). In Sect.\,\ref{specttypeO}, we noted that three stars of our sample (MSP\,167, 175 and 203) might actually be somewhat earlier than their classification listed in Table\,\ref{types}. Assuming these objects to be of type O5, rather than O6, would  increase the amount of ionizing photons by $1.5 \times 10^{49}$\,s$^{-1}$, i.e.\ by about 3\%. This latter remark does certainly not alter our conclusion about the lack of a sufficient number of ionizing photons. 

Of course, our estimate provides only a lower limit on the actual production of ionizing photons inside the RCW\,49 complex. For instance, we have not included WR\,20b in our account. This star was classified as WN7:h by Shara et al.\ (\cite{Shara}). By analogy with the ionizing fluxes of WN6ha stars in NGC\,3603 (Crowther \& Dessart \cite{CD}), the components of WR\,20a (Rauw et al.\ \cite{Rauw05}) and the models of Smith et al.\ (\cite{SNC}), we estimate that $\log(Q_0) \leq 50.0$. Thus this star alone cannot solve the issue of the missing ionizing photons. Some of the early-type stars might still be hidden from our view by the material of their natal cloud. Indeed, observations indicate that star formation is ongoing in Westerlund\,2. Whitney et al.\ (\cite{Whitney}) reported {\it Spitzer} IRAC near-IR colours of the stars in RCW\,49 that are typical of a star formation region. They estimated the total number of young stellar objects to be $\sim 7000$. Star formation occurs preferentially within $\sim 5$\arcmin\ from the cluster core, where the stellar winds of the massive stars in Westerlund\,2 have swept up gas and dust. This suggests that the ongoing star formation activity may be induced by the action of the early-type stars in Westerlund\,2. The spectral energy distributions of two stars were fitted with models of B2\,V and B5\,V stars surrounded by accretion disks and a large envelope feeding the disk (Whitney et al.\ \cite{Whitney}, Churchwell \cite{Churchwell05}). However, these stars are apparently not massive enough to contribute significantly to the ionization of the nebula. If the predictions of stellar atmosphere models are correct, then we conclude that we must still be missing an important fraction of the early-type stars in RCW\,49.\\ 

The influence of the early-type stars in the cluster core is not limited to the ionization of the nebula. The stellar winds of these stars also strongly impact on their surroundings. The signature of this process has been detected as diffuse X-ray emission with {\it Chandra} (Townsley et al.\ \cite{Townsley}). This feature is probably due to the thermalisation of the stellar winds either resulting from the collisions between the winds of individual early-type cluster members or from the shocks with the ambient interstellar medium. Townsley et al.\ (\cite{Townsley}) found that this emission has temperatures between $kT \sim 0.8$ and 3.1\,keV. Assuming a distance of 2.3\,kpc, these authors inferred a total intrinsic (i.e.\ absorption corrected) X-ray luminosity of $3 \times 10^{33}$\,erg\,s$^{-1}$ over the 0.5 -- 8.0\,keV energy range. For a distance of 8.0\,kpc, this X-ray luminosity becomes $3.6 \times 10^{34}$\,erg\,s$^{-1}$.
From the temperatures and emission measures of the best-fit X-ray models proposed by Townsley et al.\ (\cite{Townsley}), we estimate that a total thermal energy of $6.5$ -- $20.4 \times 10^{49}$\,erg (for a filling factor of 0.1 and 1.0 respectively) is required to produce the observed diffuse X-ray emission. 

Considering a Salpeter IMF and the typical masses of O-stars as tabulated by Martins et al.\ (\cite{Martins}), we estimate that Westerlund\,2 should contain about 4500\,M$_{\odot}$ in the form of stars with masses between 1 and 120\,M$_{\odot}$. We used the {\sc starburst99} code (v5.0; Leitherer et al.\ \cite{Claus}, V\'azquez \& Leitherer \cite{VL}) to estimate the total mechanical energy injected through the stellar winds. For an age of 2\,Myr, no supernova explosion should have occured. However, the total wind luminosity should be about $5.4 \times 10^{37}$\,erg\,s$^{-1}$ and the cumulative energy input from stellar winds would be about $3.6 \times 10^{51}$\,erg. This exceeds the thermal energy that produces the X-ray emission by more than one order of magnitude. Since RCW\,49 is a blister-like structure, the latter result is not surprising. Recent models accounting for the effects of clumping indicate that the mass loss rates of O-type stars might have been overestimated and are probably lower by a factor 3 -- 4 (e.g.\ Martins et al.\ \cite{clumping}, Bouret et al.\ \cite{Bouret}). Lower mass loss rates would imply a reduction of the cumulative energy input from the stellar winds, although the latter number would still exceed the thermal energy responsible for the X-ray emission by at least a factor of a few.  

In summary, while the currently known population of O-type stars in Westerlund\,2 does apparently not provide enough ionizing photons to explain the radio emission of RCW\,49, the dynamical energy provided by the stellar winds of these stars seems sufficient to explain the observed diffuse X-ray emission. 

\section{Summary and conclusions}
Our spectroscopic observations indicate that the core of Westerlund\,2 harbours at least twelve O-type stars of spectral type earlier than O7, the earliest object being of type O3\,V((f)). These spectral types are significantly earlier than previous classifications. The spectro-photometric distance of these O-type stars is found to be $8.3 \pm 1.6$\,kpc. This number is in excellent agreement with the result obtained from the study of the $B$ and $V$ light curve of the very massive Wolf-Rayet binary WR\,20a and the distance based on ZAMS fitting of the $UBV$ photometry of Moffat et al.\ (\cite{MSP}). The weighted distance of Westerlund\,2 and WR\,20a amounts to $8.0 \pm 1.4$\,kpc. This distance determination implies that the radio emission of the giant H\,{\sc ii} region RCW\,49 must be larger than previously estimated. Despite the earlier spectral types inferred in this paper, the currently known population of early-type stars in Westerlund\,2 provides only between one fifth and one third of the ionizing photons required to account for the radio emission of RCW\,49. Future observations should aim at establishing a complete census of the O-type stars population in RCW\,49, although our colour-magnitude diagram (Fig.\,\ref{coulmag}) indicates that at least within the field of view of the ANDICAM instrument, there are only a few candidates for early O-type stars other than those objects we have classified here. 

None of the early O-type stars in the cluster core displays a significant photometric variability on time scales from a day to a month and a half. None of these objects displays an obvious signature of multiplicity in our snapshot spectra. However, we have identified three eclipsing binaries (in addition to WR\,20a) that are probably late O or early B-type stars belonging to the Westerlund\,2 cluster. This suggests that there might in fact be many more (non-eclipsing) binaries. A spectroscopic follow-up of these systems as well as of some of the O-type stars that appear somewhat overluminous (MSP\,18, 183 and 203) would bring precious information on their nature as well as on the general properties of the cluster.

Finally, we emphasize that the results reported here indicate that WR\,20a actually belongs to the Westerlund\,2 cluster and further suggest that the cluster contains enough massive stars to explain the location of the Wolf-Rayet binary through the effect of a dynamical interaction scenario as suggested by Rauw et al.\ (\cite{Rauw05}).

\appendix
\section{Uncertainties of the distance determinations \label{append}}
The distance of WR\,20a stems from the luminosity determined in the light curve analysis. Under the assumptions made in Sect.\ \ref{20a}, the radii of the stars in WR\,20a are rather robust. The issue of the effective temperature is a bit trickier. This parameter is essentially determined by the spectroscopic analysis with a non-LTE model atmosphere code (Rauw et al.\ \cite{Rauw05}). The strengths of the N\,{\sc iii}, N\,{\sc iv} and N\,{\sc v} features provide the most stringent constraints. The model atmosphere code used by Rauw et al.\ (\cite{Rauw05}) assumes a spherically symmetric star. In a binary system such as WR\,20a, the stars are however not spherical. Their shape is rather set by the Roche potential. Moreover, the gravity varies over the stellar surface, being maximum near the poles and minimum on the portion of the surface facing the companion star. As a result, gravity darkening renders the surface temperature non-uniform: the star is hotter near the poles and cooler near the equator. Adopting the fitting parameters of the light curve shown in Fig.\, \ref{lc20a} and assuming a gravity darkening exponent $\beta = 1.0$, we estimate that the temperature near the primary's poles should be about 1900\,K hotter than the average effective temperature (43000\,K), while it would be lower by $\sim 5300$\,K over the part of the stellar surface facing the secondary. Whilst these numbers may be overestimated, for instance because we neglected reflection effects, they nevertheless suggest that part of the difficulties encountered when trying to fit simultaneously N\,{\sc iii}, N\,{\sc iv} and N\,{\sc v} lines might actually arise from a non-uniform surface temperature distribution in a close binary system. Concerning the distance of WR\,20a, we estimate that the effect of gravity darkening could produce a 4 -- 5\% uncertainty on the effective temperature which would then correspond to a 10\% margin on the distance.

Alternatively, our distance estimate of the early O-type stars in Westerlund\,2 could be biased. In this respect, we note the rather large dispersion ($0.65$\,mag) of our spectro-photometric distance moduli. The primary suspects to bias our result would be binary systems. However, our photometric monitoring campaign failed to reveal any eclipsing binary system among these O-type stars with a period between a few days and a few weeks. Furthermore, we do not find evidence for multiplicity in our spectroscopic data either, although these snapshot data are certainly not suited to provide firm constraints on the number of spectroscopic binaries. Therefore, it seems unlikely that the majority of the O-type stars that we have observed are binaries. If we nevertheless consider that some of them might be multiple objects, the smaller distance moduli of MSP\,18, 183 and 203 make them the best binary candidates. Discarding these three objects from the mean yields $DM = 14.72 \pm 0.54$ (1-$\sigma$ standard deviation). Note that the dispersion about the mean remains quite large, though comparable to the cosmic scatter of absolute magnitudes based on the spectral types. 

Another possible reason for the dispersion of $DM$ could be that some of the main-sequence O-stars are slightly evolved off the main sequence and are thus intrinsically brighter than `typical' main-sequence stars. Our luminosity class determination based on the strength of the He\,{\sc ii} $\lambda$\,4686 line suggests that some of the stars are indeed giants. Those stars that have an $EW$ of the He\,{\sc ii} $\lambda$\,4686 line at the border-line between the two luminosity classes are labelled with a V-III luminosity class in Table\,\ref{types}. For these objects, we evaluated the distance by comparison with an interpolation of the main-sequence and giant absolute magnitudes of Martins \& Plez (\cite{MP}). Some other objects could have intermediate evolutionary stages and might thus be responsible for part of the dispersion. 

Finally, as suggested in Sect.\,\ref{latety}, some part of the reddening that affects the stars in Westerlund\,2 might be produced by material from the parental molecular cloud. Therefore, we cannot rule out a priori that, at least for this material, $R_V$ differs from the canonical value of 3.1. In fact, in some clusters with a higher mean density interstellar medium, it has been argued that $R_V$ could range from 3 to 6 (see e.g.\ Turner \cite{Turner}). If that were the case for Westerlund\,2, all the distance determinations in this paper would have to be revised downwards. To determine the value of $R_V$ requires at least $UBV$ photometry. Since we do not have our own new $U$-band measurements, we would have to rely on the data of Moffat et al.\ (\cite{MSP}) or Carraro \& Munari (\cite{CM}). Unfortunately, both data sets seem to have problems (see Sect.\,\ref{photom}) and their $U$ magnitudes disagree. Our attempts to use these archive photometric data failed to produce conclusive results and we have to conclude that the reddening law cannot be quantified with the data currently available.

\acknowledgement{We are grateful to the collaborators of the SMARTS consortium for obtaining the data of our photometric monitoring campaign. We thank Chang-Hui Rosie Chen (UIUC) for providing us with the {\sc starburst99} simulations, Prof.\ David Turner (Saint Mary's University) for discussion about the reddening law and the anonymous referee for his/her comments that helped us improve our manuscript. The Li\`ege group acknowledges financial support from the FNRS (Belgium), as well as through the XMM and INTEGRAL PRODEX contract and contract P5/36 PAI (Belspo). AFJM acknowledges financial support from NSERC (Canada) and FQRNT (Qu\'ebec).}

%\listofobjects
\end{document}